\documentclass[11pt]{article}
\usepackage{graphics}
\usepackage{amsmath,amsthm,amssymb}
\newcommand{\be}{\begin{equation}}
\newcommand{\ee}{\end{equation}}
\newcommand{\bea}{\begin{eqnarray}}
\newcommand{\eea}{\end{eqnarray}}

\usepackage[pdftex]{graphicx}

\usepackage{harvard}

\title{\bf A Robust Bayesian Dynamic Linear Model for Latin-American Economic Time Series:
``The Mexico and Puerto Rico Cases"}

\author{Jairo F\'uquene$^{a}$ \thanks{Corresponding author.},  Marta \'Alvarez.$^{b}$
and Luis Ra\'ul Pericchi$^{c}$.\\
\\ \small $^{a}$ Department of Statistics. University of Warwick. UK.
J.A.Fuquene-Patino@warwick.ac.uk.
\\ \small $^{b}$ Institute of Statistics,
Business School, University of Puerto Rico, Rio Piedras.
\\ \small $^{c}$ Department of Mathematics. University of Puerto Rico, Rio Piedras.
}

\begin{document}

\maketitle

\bibliographystyle{agsm}
\citationstyle{agsm}


\begin{abstract}

The traditional time series methodology requires
at least a preliminary transformation of the data to get
stationarity. On the other hand, Robust Bayesian Dynamic Models
(RBDMs) do not assume a regular pattern or stability of the
underlying system but can include points of statement breaks.
In this paper we use RBDMs in order to account possible outliers and structural breaks
in Latin-American economic time series. We work with important economic time series from Puerto Rico and Mexico.
We show by using a random walk model how RBDMs can be applied
for detecting historic changes in the economic inflation of Mexico. Also,
we model the Consumer Price Index (CPI), the Economic
Activity Index (EAI) and the total number of employments (TNE) economic time series in Puerto Rico using local linear trend and seasonal RBDMs
with observational and states variances. The results illustrate how the model
accounts the structural breaks for the historic recession periods in Puerto Rico.

\end{abstract}

\textbf{Keywords:} {\it Robust Bayesian Dynamic Model, Outliers and Structural Breaks,
Latin-American Time Series, Consumer Price Index,  Economic
Activity Index, Total Number of Employments.}\\\\

\textbf{JEL Classification: C11 $\cdot$ C40 $\cdot$ G17 $\cdot$ N16}

\section{Introduction}

Economic  Latin-American time series variables can be complex with high
frequency data. From a frequentist perspective, techniques for fitting time series models
use a preliminary data-set 

\begin{figure}[h]

transformation in order to get
stationarity and therefore important information about the dynamic system can be lost.
On the other hand, from a  Bayesian perspective
the use of  RBDMs with weakly-robust priors for the observation and state variances has been revolutionary in recent years due to flexibility for detecting outliers and structural breaks, the straightforward computational techniques based in Markov Chain Monte Carlo (MCMC)
and the natural update from Bayes theorem without a preliminary data-set transformation.\\
From a frequentist perspective the study of time series
with structural changes has been of far reaching in econometric theory
for univariate time series, frequentist dynamic models, volatility
and even financial return models.
The seminal paper of \citeasnoun{Tsay} considers least square techniques
and residual variance ratios for detecting outliers, level shifts and variance changes in univariate time series.
\citeasnoun{Hansen} introduces a bootstrap method for detecting
structural changes in regressors including structural shifts, polynomial
trends and exogenous stochastic trends
for frequentist dynamic econometric models.
For volatility models, \citeasnoun{Vladimir} uses local chance
point analysis to intervals of homogeneity in order to account possible structural breaks. \citeasnoun{Fryzlewicz}
propose a method based in process transformation and binary segmentation
for detecting multiple change points
in auto-regressive conditional heteroscedastic models for financial returns.\\
From a Bayesian perspective in recent years RBDMs
for detecting structural breaks have been proposed as an alternative
to the usual Bayesian dynamic models. \citeasnoun{Ardia}
propose a Bayesian generalized autoregressive conditional heteroscedasticity dynamic model with Student-t innovations with applications
to the R program (\citeasnoun{RR}).
\citeasnoun{polsong} apply heavy-tailed priors in order to examine
historical patterns of return on assets to financial time series. \citeasnoun{fuque2} propose
a new flexible class of heavy-tailed priors for detecting outliers
and structural breaks in Bayesian dynamic linear models.\\
However, even though the qualities of RBDMs in the best of our knowledge
there are no
application of RBDMs to economic Latin-American variables for detecting their historic outliers and
structural breaks. Therefore, in this
work we use the RBDMs proposed by  \citeasnoun{fuque2} to modelling
the historic change points in  Latin-American economic variables
from Mexico  and Puerto Rico. We use this methodology
because: 1) We can model a considerable variety of dynamic models: random walk, local linear trend and seasonal
or a combination of those models with weakly-robust
priors for the observation and state variances, 2) The computational schemes can be applied easily  for practitioners and
3) Using RBDMs allow us to have posterior inference for the parameters
from a Bayesian perspective. \\
The paper is organized as follows. Section 2 shows the prior
variances specification for the RBDM. In Section 3, we
apply a random walk RBDM to an economic time series of Mexico for
detecting the historic outliers and level changes in the inflation
of this country.  Section 4 shows local trend and stationary RBDMs
for accounting the abrupt changes in the economic recession periods in Puerto Rico.
Finally we have the conclusions in Section 5.

\section{Model Specification}

The Dynamic Linear Model (DLM) is specified (see
\citeasnoun{west2}) by a Normal prior distribution for the $p$-dimensional
state vector at $t=0$ as follows:
\begin{align}
\boldsymbol{\theta}_{0}\sim N_{p}(m_{0},C_{0}),
\end{align}

\end{figure}

\clearpage

\begin{figure}[h]

with the set of equations:
\begin{align}
y_{t}&=F_{t}\boldsymbol{\theta}_{t}+\nu_{t} \;\;\;\; \nu_{t}\sim N_{m}(0,V_{t}),\\
\boldsymbol{\theta}_{t}&=G_{t}\boldsymbol{\theta}_{t-1}+\omega_{t}\;\;\;\;
\omega_{t}\sim N_{p}(0,W_{t}),
\end{align}
with $t=1:T$ and where $F_{t}$ and $G_{t}$ are known matrices
of order $p \times p$ and $m \times p$ respectively. With $\nu_{t}$ and
$\omega_{t}$ two independent Gaussian random vectors with mean
zero and known variance $V_{t}$ and $W_{t}$ respectively. The observation equation and
state equation are (2) and (3), respectively. A set of prior distributions
for the observation and state variances may be considered in practice. For example
in order to have closed form full conditionals we could use gamma prior densities. However, in the presence
of highly frequency data heavy-tailed priors are the best alternative. The scaled Beta2
prior for the precision $\lambda=1/\tau^{2}$ is proposed in \citeasnoun{fuque2}
for modelling the variances (and precisions) in DLMs and defined as follows:

\begin{equation}
\pi(\lambda)=\frac{\Gamma(q+p)}{\Gamma(q)\Gamma(p)}\beta\frac{(\beta\lambda)^{q-1}}{\left(1+\beta\lambda\right)^{p+q}};\;\;\;\;
\lambda>0
\end{equation}

where $\beta$ is the scale parameter. This paper consider the
Student-t density coupled with a scaled Beta2
for modelling the observation and state errors (as in \citeasnoun{fuque2}) in Latin-American
economic time series from Mexico and Puerto Rico. So, let $\theta\sim$ be a
Student-t$(\mu,\tau,\upsilon)$ where $\upsilon$ are the degrees of
freedom, $\mu$ the location and $\tau$ the scale of the Student-t
density:
\begin{equation}
\pi(\theta|\tau^{2})=\frac{k_{1}}{\tau}\left(1+\frac{1}{\upsilon}
\left(\frac{\theta-\mu}{\tau}\right)^{2}\right)^{-(\upsilon+1)/2},
\;\;\;\; \upsilon>0, -\infty<\mu<\infty, -\infty<\theta<\infty,
\end{equation}
where
$k_{1}=\dfrac{\Gamma((\upsilon+1)/2)}{\Gamma(\upsilon/2)\sqrt{v\pi}}$.
We have that
$\pi(\theta)=\int_{0}^{\infty}\pi(\theta|\tau^{2})\pi(\tau^{2})d\tau^{2}$
and therefore the marginal prior as follows:
\begin{align*}
\pi(\theta)=
\begin{cases}
  \beta^{q}\nu/(\theta-\mu)^{q+1/2}2F1(p+q,q+1/2,(\upsilon+1)/2+p+q,1-\beta\nu/(\theta-\mu)^{2}) & \text{if\;\;\;} \theta\neq\mu, \\
\\
   k_{1}\text{Be}(q+1/2,p+v/2)/\text{Be}(p,q) & \text{if\;\;\;} \theta=\mu,
 \end{cases}
\end{align*}
with $2F1(a,b,c,z)$ the hypergeometric function (see 15.1.1 of
\citeasnoun{book_integra}) and we have that $\pi(\theta)$ is the
Student-t-Beta($\upsilon$,p,q,$\beta$) (see \citeasnoun{fuque2}
for the proof of this result). The variances of the RBDM are
Student-t-Beta($\upsilon$,q,p,$\frac{1}{\beta}$) densities (with the Beta2
prior for the precision as $\lambda=1/\tau^{2}$). Here $W_{t,i}$ denotes
the $i$th diagonal element of $W_{t,i}$, $i=1,...,n$ the
hierarchical Student-t-Beta($\upsilon$,q,p,$\frac{1}{\beta}$)
prior can be summarized as follows:

\begin{align*}
\hspace{1cm} V_{t}^{-1}=&\lambda_{y}\omega_{y,t}, & W_{t,i}^{-1}=&\lambda_{\theta,i}\omega_{\theta,t_{i}},\\
\lambda_{y}|q& \sim \text{Gamma}(q,(\beta\rho_{y})^{-1}), &
\lambda_{\theta,i}|q& \sim \text{Gamma}(q,(\beta\rho_{\theta,t_{i}})^{-1}),\\
\omega_{y,t}& \sim \text{Gamma}(\upsilon/2,2/\upsilon), &
\omega_{\theta,t_{i}}&
\sim \text{Gamma}(\upsilon/2,2/\upsilon),\\
\rho_{y}&\sim \text{Gamma}(p,1), & \rho_{\theta,t_{i}} &\sim
\text{Gamma}(p,1),
\end{align*}

For each $t$, the posterior distribution of the latent variables
$\omega_{y,t}$ and $\omega_{\theta,t_{i}}$ is useful in order to account the
outliers and abrupt changes in the economic time series. Values of $\omega_{y,t}$ and
$\omega_{\theta,t_{i}}$ smaller than one indicate possible
outliers or abrupt changes respectively. A Gibbs sampler scheme can be implemented by using the
full conditional in closed form of RBDMs (see Appendix A).

\end{figure}

\clearpage

\begin{figure}[h]

\subsection{Illustration RBDM with a toy example : the annual CPI from Puerto Rico}

We consider now the annual Consumer Price Index in Puerto in the log-scale in order
to illustrate how a RBDM works. We use  a local linear trend model (i.e., linear growth
model) for fitting the trend and slope of the CPI in logarithm scale as follows:
\begin{align}
y_{t}&=\mu_{t}+\nu_{t}, \;\;\;\; &\nu_{t}\sim N(0,V_{t}),\notag\\
\mu_{t}&=\mu_{t-1}+\xi_{t-1}+\omega_{t,1}, \;\;\;\; &\omega_{t,1}\sim N(0,W_{t,1}), \\
\xi_{t}&=\xi_{t-1}+\omega_{t,2}, \;\;\;\; &\omega_{t,2}\sim \notag
N(0,W_{t,2}), \notag
\end{align}
with uncorrelated errors $\nu_{t}$, $\omega_{t,1}$ and
$\omega_{t,2}$ and where

\begin{align*}
\theta_{t}&=\left[
\begin{array}{c}
\mu_{t}\\
\xi_{t}
\end{array}\right],&
G&=\left[
\begin{array}{cc}
1 & 1\\
0 & 1
\end{array}\right],&
W_{t}&=\left[
\begin{array}{cc}
\sigma^{2}_{\mu,t} & 0\\
0 & \sigma^{2}_{\xi,t}
\end{array}\right],&
F=&\left[\begin{array}{cc}
 1 & 0
\end{array}\right].
\end{align*}

\begin{center}\vspace{3.7cm} \hspace{-4.5cm}
\includegraphics[bb=0 0 10cm 10cm,scale=0.7,keepaspectratio]{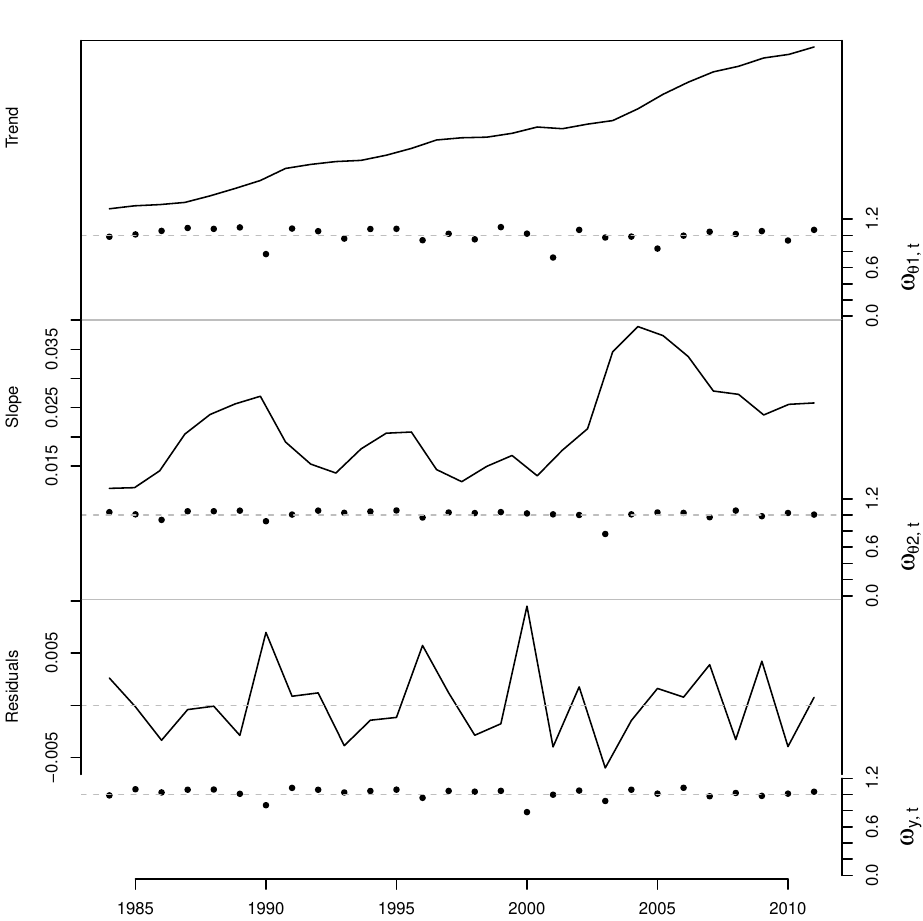}
\caption{Outliers and structural breaks in the annual logarithm
Consumer Price Index in Puerto Rico using the robust approach. The right scale
is for the latent $\omega_{y,t}$ and $\omega_{\theta,t_{i}}$ parameters.}
\end{center}

For this toy example we use a Student-t-Beta2 where $p=q=1$, and
$1/\beta=10000$ as is proposed in \citeasnoun{pericchi1}. We
have convergence of all parameters using 10000 iterations
after a burn-in phase period of 5000 iterations. Figure a displays
how by using the RBDM an outlier in the year 2000 is obtained. The changes in the
trend shows the level changes in 1990 and 2001 and the slope presents a change in
2003.

\end{figure}

\clearpage

\begin{figure}

\section{The case of Mexico}

In this section we use a random walk RBDM
in order to detect outliers and structural breaks
of the inflation in Mexico. We use the monthly logarithm of the CPI-variations from
1696 to 1983 in order to accounting possible changes in inflation in
Mexico during this period such as: 1) The monetary devaluation
in 1976 and 1982. 2) The value-added tax imposed in 1980 with the
posterior modification in 1983. 3) Some changes in the payment to employees and
the increase in gasoline prices and
4) The modification in the economy in Mexico in 1983. The random walk model RBDM can be written as follows:

\begin{align}
y_{t}&=\theta_{t}+\nu_{t}, \;\;\;\; &\nu_{t}\sim N(0,V_{t}),\notag\\
\theta_{t}&=\theta_{t-1}+\omega_{t}, \;\;\;\; &\omega_{t,1}\sim N(0,W_{t}),
\end{align}

with the prior specification showed in previous Section. We implement the Gibbs sampling scheme showed
in Appendix A and from a visual assessment of the Gibbs output
we have that convergence has been achieved in Appendix B. The ergodic means are nonetheless pretty stable in the middle
of the plots and the decay of the empirical autocorrelation function is very fast.

\begin{center}\vspace{3.5cm} \hspace{-6.0cm}
\includegraphics[bb=0 0 10cm 10cm,scale=0.75,keepaspectratio]{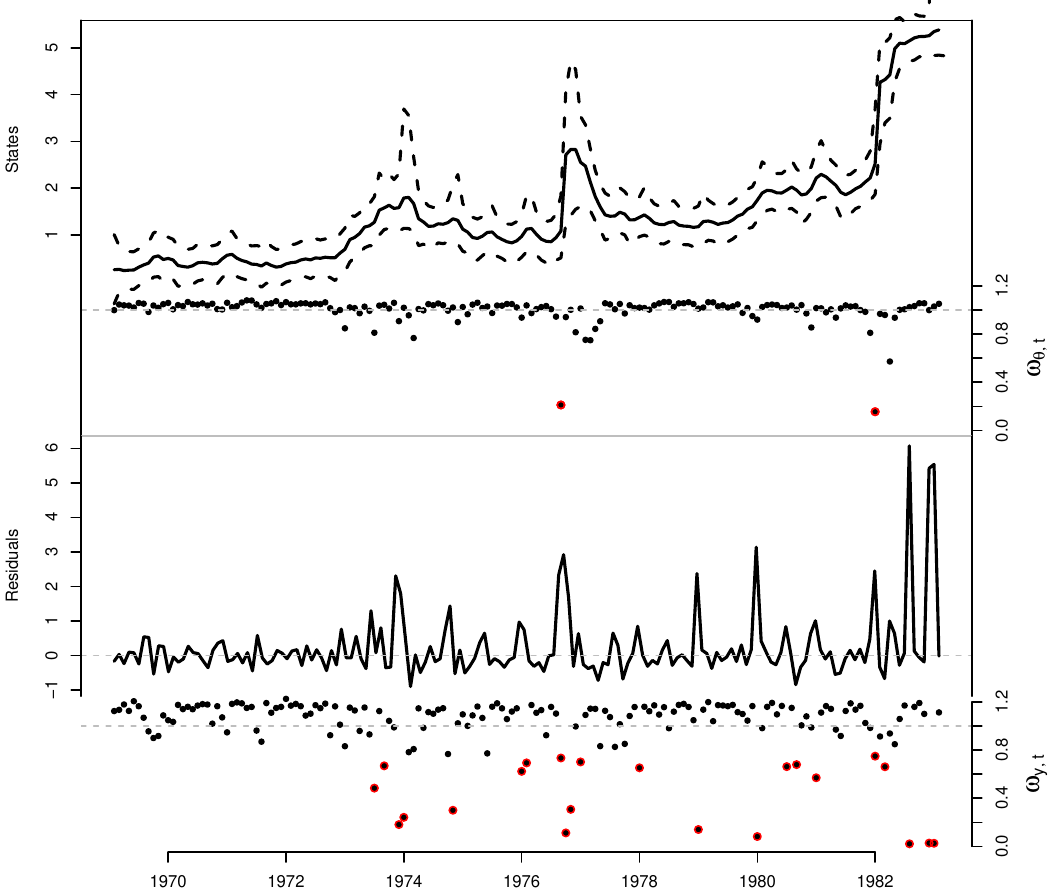}
\caption{Outliers and structural breaks in CPI-variations from Mexico 1969-1983. The right scale
is for the latent $\omega_{y,t}$ and $\omega_{\theta,t_{i}}$ parameters. Red points illustrate the outliers and structural breaks.}
\end{center}

\end{figure}

\clearpage

\begin{table}[htb]

\scriptsize { \caption{Posterior mean of $\omega_{y,t}$
for the monthly CPI-variations from Mexico 1969-1983.}\label{simula2}
\renewcommand{\arraystretch}{1.4}
\begin{center}
\begin{tabular}{cccccccc}\\\hline
\hline
month/year & $E(\omega_{y,t}|y_{1:T})$ \\\hline
\hline
Jul 1973 & 0.8118393\\
Dec 1973 & 0.18126535 \\
Jan 1974 & 0.24155247  \\
Nov 1974 & 0.30013551  \\
Oct 1976 & 0.11182539  \\
Nov 1976 & 0.30769569  \\
Jan 1979 & 0.14002184  \\
Jan 1980 & 0.08193927  \\
Aug 1982 & 0.02124920  \\
Dec 1982 & 0.02747282   \\
Jan 1983 & 0.02584617  \\
\hline
\end{tabular}
\end{center}
}

\end{table}

According to Figure 2. the state parameters show significant level changes
in September 1976  and January 1982. These changes likely represent the exchange rate devaluation
in September 1976 and just one month before of the second devaluation in 1982. On the other hand,
in Table 1 the expectations for the latent parameters for identifying the outliers
in the random walk RBDM are presented. The outliers in January 1980 and
January 1983 probably showed the value-added tax imposed and consequently modified
in those dates with also the modification in the economy in Mexico in 1983.
The most dramatic increase in payment to employees could be
exposed for the extreme values in January 1974, October 1976 and
January 1979. The exchange rate devaluation presented in  August 1982 and December 1982
are also presented in Table 1. Due to the increase in gasoline prices
the RBDM could show outliers in December 1973, November 1974 and
November 1976. Finally, we think this random walk RBDM  could be useful not only to accounting changes
in this economic time series but also may be useful as an pre-intervention dynamic model.

\section{The case of Puerto Rico}

In this section we use the linear trend RBDM presented in (6) and a linear trend RBDM
with a seasonal component in order to model the logarithm of
CPI, AEI and TNE from January 1980 to
December 2012 in Puerto Rico. The CPI and EAI are two economic indexes widely
used for describing the economic situation of Puerto Rico. The CPI and EAI
are useful for accounting the inflation through of price fluctuations and
the real economic activity. EAI and TNE are very correlated in the sense
that EAI is computed by using also TNE. However TNE is an interesting
time series for the quarterly seasonal component and also
for the historic fact that in Puerto Rico in July 2009 near of 17000
employments lost their jobs for the recent economic crisis. We find that by using RBDMs
historical changes are detected as structural breaks in the trend of the models.

\clearpage

\begin{figure}[h] 

Now we describe briefly some of the important historical changes useful
for the interpretation of the results.\\
\textbf{CPI historical changes.}
A first index for accounting the inflation in Puerto Rico
was born by using the cost of living for working families in 1940.
The CPI was born in 1977 by including in the first index
information on urban
families, self-employed and the pensioners in Puerto Rico. Using
addition products, in 1990 a few adjustments to
items and services of
CPI were proposed. A new study of Income and Expenses was made
in the years 1999-2003 with a major change in the CPI methodology
in March 2010. Currently, the basket of goods
for CPI has the following major groups: food
and beverages, housing, apparel, transportation, medical care,
entertainment, education and communication and
those groups are similar to the United
States basket  (see for example \citeasnoun{uno}).\\
\textbf{EAI historical changes.} The monthly  EAI includes the
behavior of four economic indicators: total number of employments
(thousands), cement sales (million bags), fuel consumption
(millions of gallons) and electricity generation (million KWH).
According to the  \citeasnoun{dos} the EAI has a strong
linear correlation of 0.97 with the Gross National Product (GNP). So, in this work we use EAI
as one of the indicators useful for detecting recession periods
in the economy of Puerto Rico during the last 35 years have been: 1980-82, 1990-91, 2001-02 and since
2006.\\

\begin{center}
\hspace{-7.5cm}
\includegraphics[bb=0 0 10cm 10cm,scale=0.5,keepaspectratio]{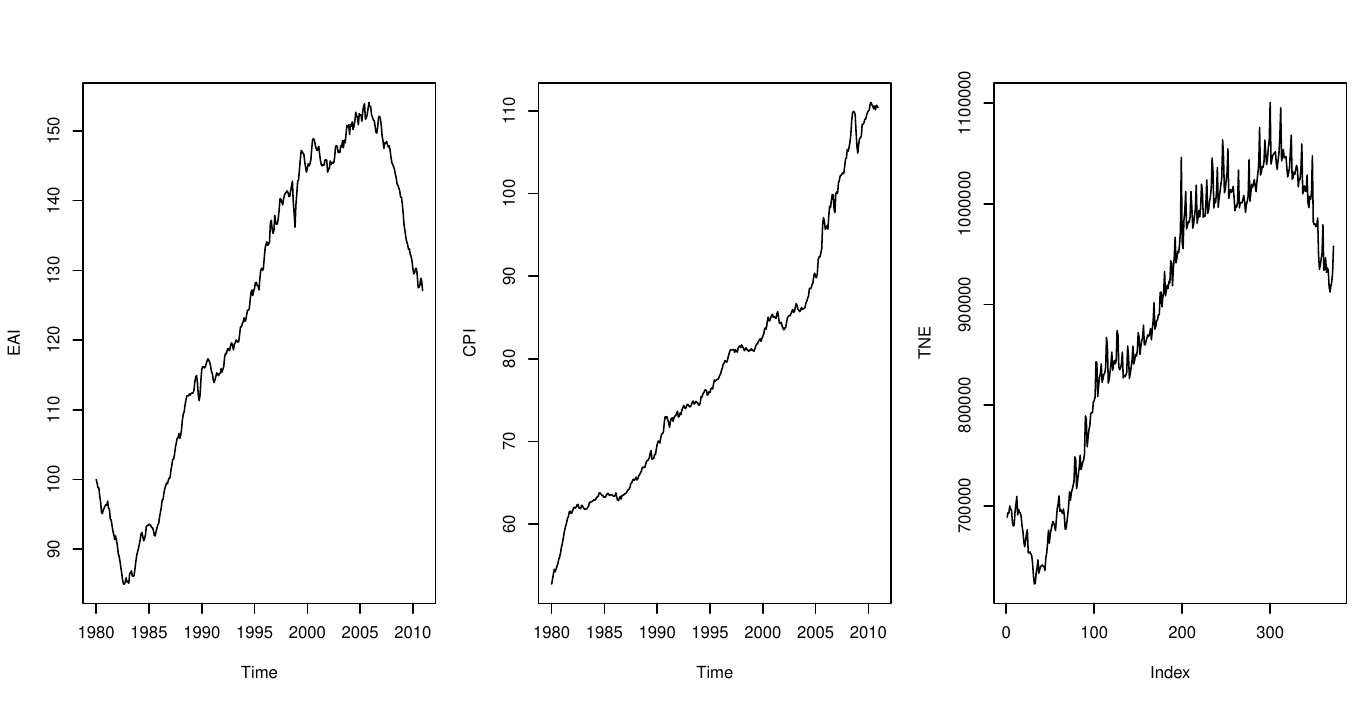}
\caption{Economic time series from Puerto Rico: EAI, CPI and TNE.}
\hspace{-9.5cm}
\end{center}

Due to the quarterly seasonal component in TNE (see Figure 3)
we use a local trend with a seasonal component in this economic variable. The observation and system matrices of this
model are:
\begin{align*} F&=\left[ \begin{array}{ccccc} 1 & 0 & 1 & 0 & 0 \end{array}\right] & G&=\left[\begin{array}{ccccc} 1 &
1 & 0 & 0 & 0\\ 0 & 1 & 0 & 0 & 0\\ 0 & 0 & -1 & -1& -1\\ 0 & 0 & 1 & 0 & 0\\ 0 & 0 & 0 & 1 & 0 \end{array}\right],
\end{align*}
and the unknown parameters are the observations variance $V_{t}$ and three elements for $W_{t}$:
\begin{align*} W_{t}=\left[ \begin{array}{ccccc} \sigma^{2}_{\mu,t}, & \sigma^{2}_{\xi,t}, & \sigma^{2}_{s,t}, & 0, & 0
\end{array} \right] \end{align*}
where $\sigma^{2}_{\mu,t}$, $\sigma^{2}_{\xi,t}$ and $\sigma^{2}_{s,t}$ are the unknown variances of the level of the
series, the slope of the linear trend and the seasonal respectively.

\end{figure}

\clearpage

\begin{figure}[h] 

\begin{center}\vspace{3.5cm} \hspace{-6.0cm}
\includegraphics[bb=0 0 10cm 10cm,scale=0.6,keepaspectratio]{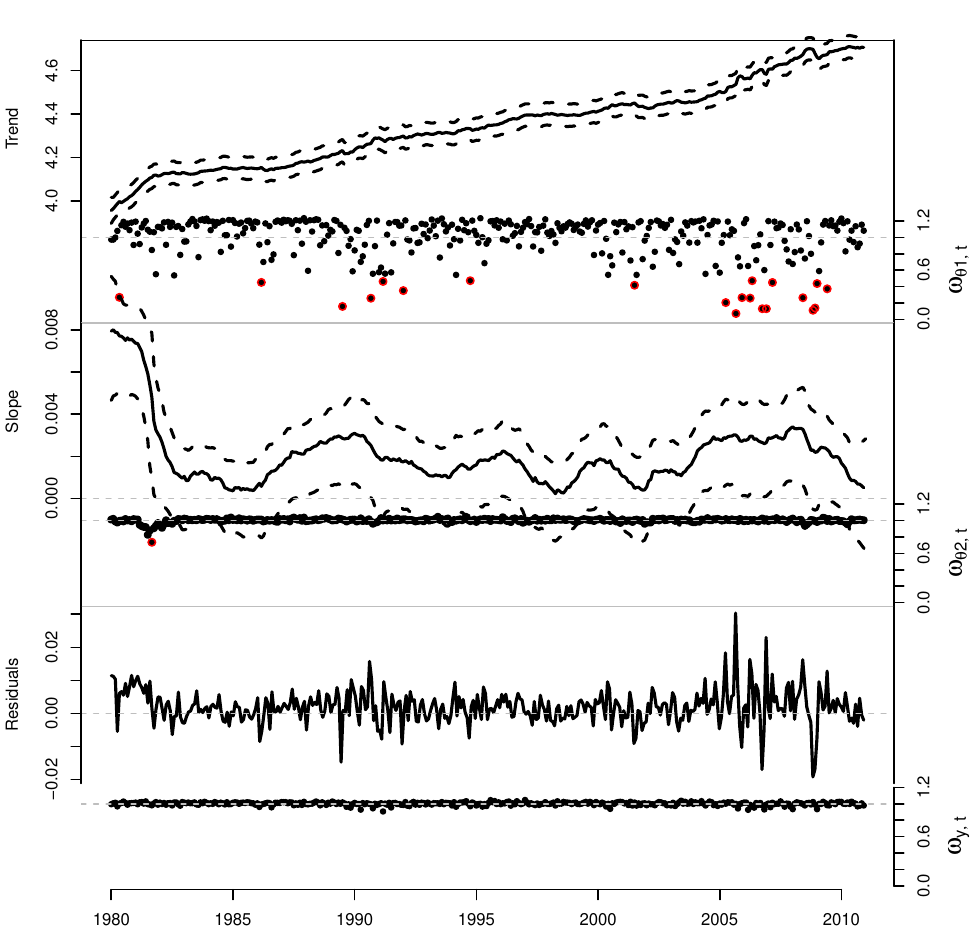}
\caption{Outliers and structural
breaks in the logarithm monthly Consumer Price Index in Puerto
Rico. The right scale
is for the latent $\omega_{y,t}$ and $\omega_{\theta,t_{i}}$ parameters. Red points illustrate the outliers and structural breaks.}
\end{center}
\begin{center}\vspace{3.5cm} \hspace{-6.0cm}
\includegraphics[bb=0 0 10cm 10cm,scale=0.6,keepaspectratio]{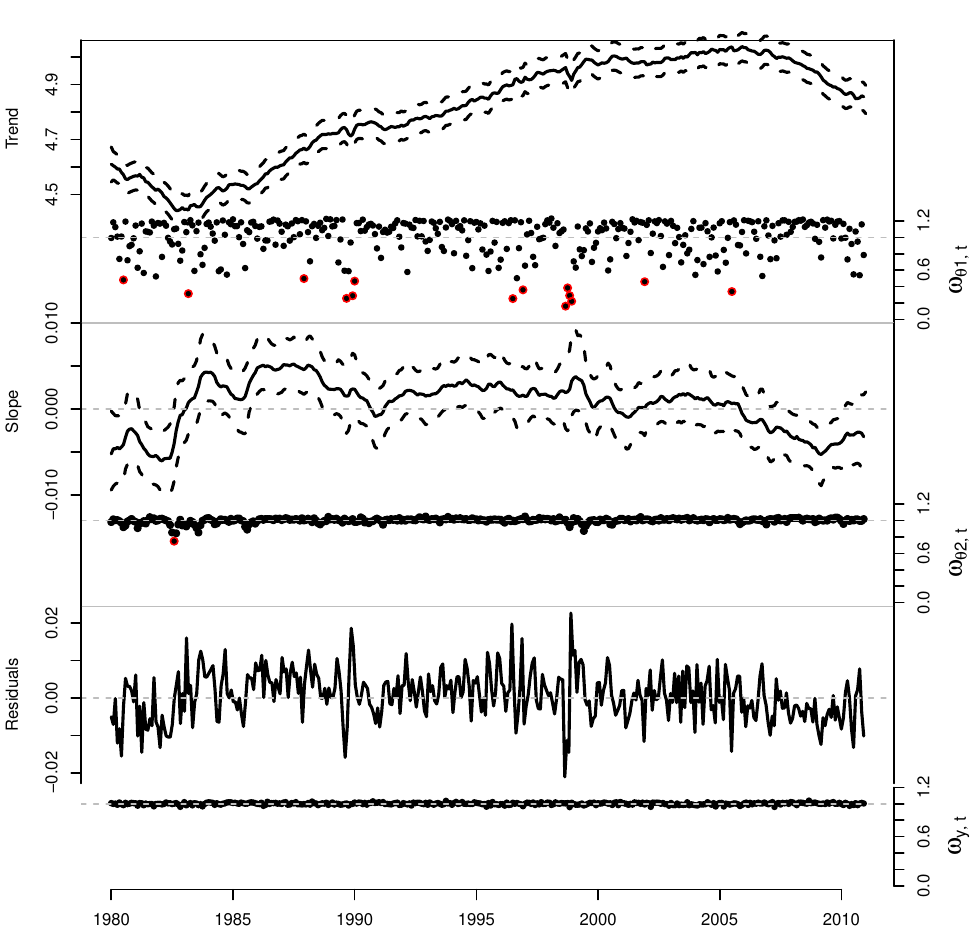}
\caption{Outliers and structural breaks in the logarithm monthly
Economic Activity Index in Puerto Rico. The right scale
is for the latent $\omega_{y,t}$ and $\omega_{\theta,t_{i}}$ parameters. Red points illustrate the outliers and structural breaks.}
\end{center}

\end{figure}

\clearpage

\begin{figure}[h] 

Figure 4 displays the results by using CPI
for the period of January 1980 to December 2012. The residuals in bottom of Figure 4 are given by
$\hat{\epsilon}_{t}=y_{t}-E(F\theta_{t}|y_{1:T})$. By looking at
the residuals there are no outliers.
The slope is dynamically changing with only a sudden jump
within the first recession period in September 1981.

\begin{center}\vspace{4.5cm} \hspace{-6.0cm}
\includegraphics[bb=0 0 10cm 10cm,scale=0.85,keepaspectratio]{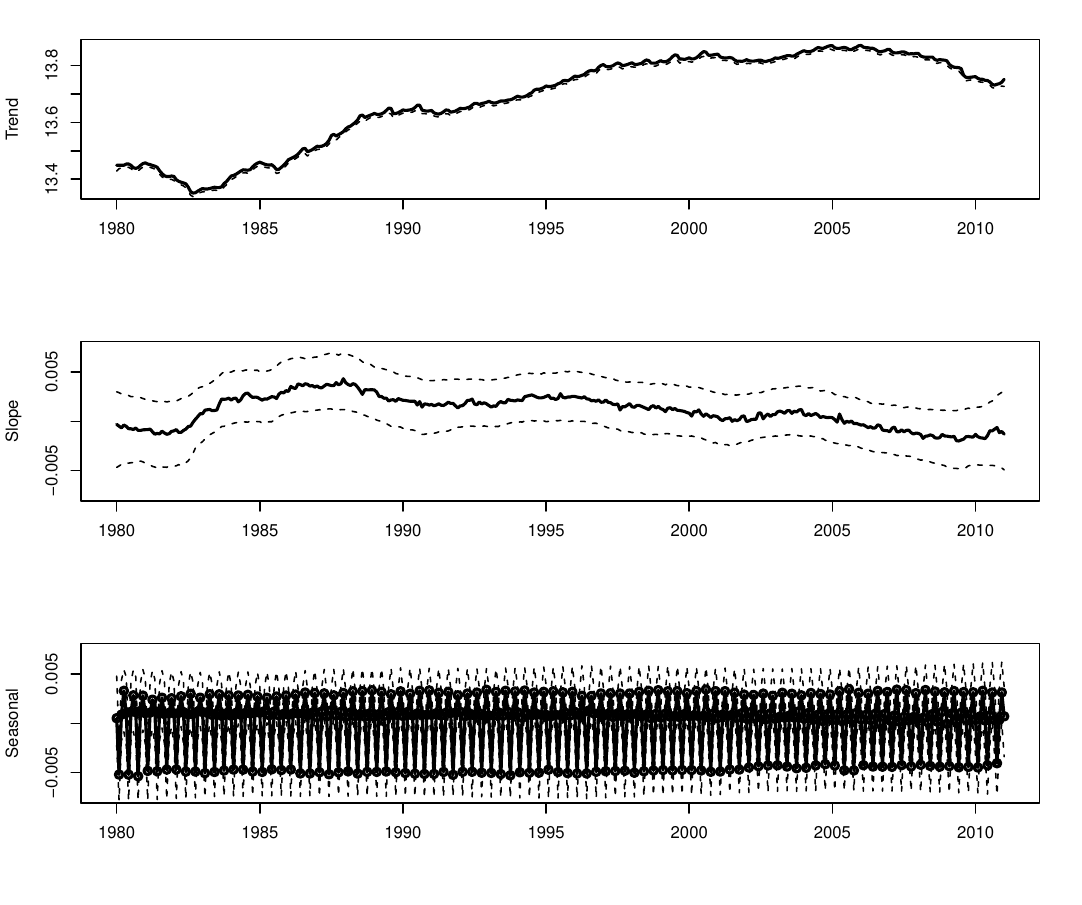}
\caption{Trend, slope and seasonal posterior mean parameters with their corresponding
credible interval for TNE.}
\end{center}

Table 2 show how the trend has different jumps, the most dramatic one in September
2005 with $E(\omega_{\theta,t_{1}}|y_{1:T})=0.07$  and some other
abrupt changes in the precedent years. Even though the Consumer
Price Index is an indicator of inflation, this
dramatic change could have been an ``alarm" for the economic
recession Puerto Rico is facing since 2006. Other dramatic changes
are found in May 1980, July 1989 and September 1990, all of them
within of recession periods in Puerto Rico.

\end{figure}

\clearpage

\begin{table}[htb]

The structural break in 1980 likely is due to changes in
the methodology for computing the CPI. The change in the trend of the CPI at the years 1989
and 1990 could also be related to the implementation of  the
``Joint Committee on Taxation" in the United States, and his
effect on the island. Figure 5 shows the abrupt changes in
the trend for the EAI.

\scriptsize { \caption{Posterior mean of $\omega_{\theta,t_{1}}$
for the monthly logarithm of the CPI, EAI and TNE.}\label{simula2}
\renewcommand{\arraystretch}{1.4}
\begin{center}
\begin{tabular}{cccccccc}\\\hline
\hline
month/year & $E(\omega_{\theta,t_{1}}|y_{1:T})$ - CPI \\
\hline
May 1980 & 0.26805231 \\
Mar 1986 & 0.45021372 \\
Jul 1989 & 0.15716954 \\
Sep 1990 & 0.25732717 \\
Mar 1991 & 0.46168657 \\
Jan 1992 & 0.35024676 \\
Oct 1994 & 0.47112292 \\
Jul 2001 & 0.41570839 \\
Apr 2005 & 0.20405587 \\
Sep 2005 & 0.07041677 \\
Dec 2005 & 0.26448903 \\
Apr 2006 & 0.25815601 \\
May 2006 & 0.47095720 \\
Dec 2006 & 0.12863052 \\
Jun 2008 & 0.26322651 \\
Nov 2008 & 0.10991814 \\
Dec 2008 & 0.13796592 \\
Jan 2009 & 0.43613370 \\
Jun 2009 & 0.37176951 \\
\hline
\end{tabular}
\end{center}
}
\renewcommand{\arraystretch}{1.4}
\begin{center}
\begin{tabular}{cccccccc}\\\hline
\hline
month/year & $E(\omega_{\theta,t_{1}}|y_{1:T})$ - EAI \\
 \hline
Jul 1980 & 0.4805507 \\
Mar 1983 & 0.3128145 \\
Dec 1987 & 0.4969033 \\
Sep 1989 & 0.2555551 \\
Dec 1989 & 0.2868322 \\
Jan 1990 & 0.4662116 \\
Jul 1996 & 0.2527494 \\
Dec 1996 & 0.3596205 \\
Sep 1998 & 0.1603856 \\
Oct 1998 & 0.3832146 \\
Nov 1998 & 0.2897125 \\
Dec 1998 & 0.2202965 \\
Dec 2001 & 0.4586847 \\
Jul 2005 & 0.3384897 \\\hline
\end{tabular}
\end{center}
\renewcommand{\arraystretch}{1.4}
\begin{center}
\begin{tabular}{cccccccc}\\\hline
\hline
month/year & $E(\omega_{\theta,t_{1}}|y_{1:T})$ - TNE.\\
 \hline
Aug 1989 & 0.4428376 \\
Aug 1990 & 0.4871985 \\
Jul 2009 & 0.4016160\\\hline
\end{tabular}
\end{center}

\end{table}

\clearpage

\begin{figure}[h]

In particular, two dramatic level changes are presented in
September 1989 and December 1989 in the beginning of the
recession period of 1990. On the other hand other structural breaks are
showed in July 1996, September 1998, December
1998 and July 2005 probably related with the rest of
recession periods.

\begin{center}
\vspace{1.5cm} \hspace{-9.5cm}
\includegraphics[bb=0 0 12cm 12cm,scale=0.7,keepaspectratio]{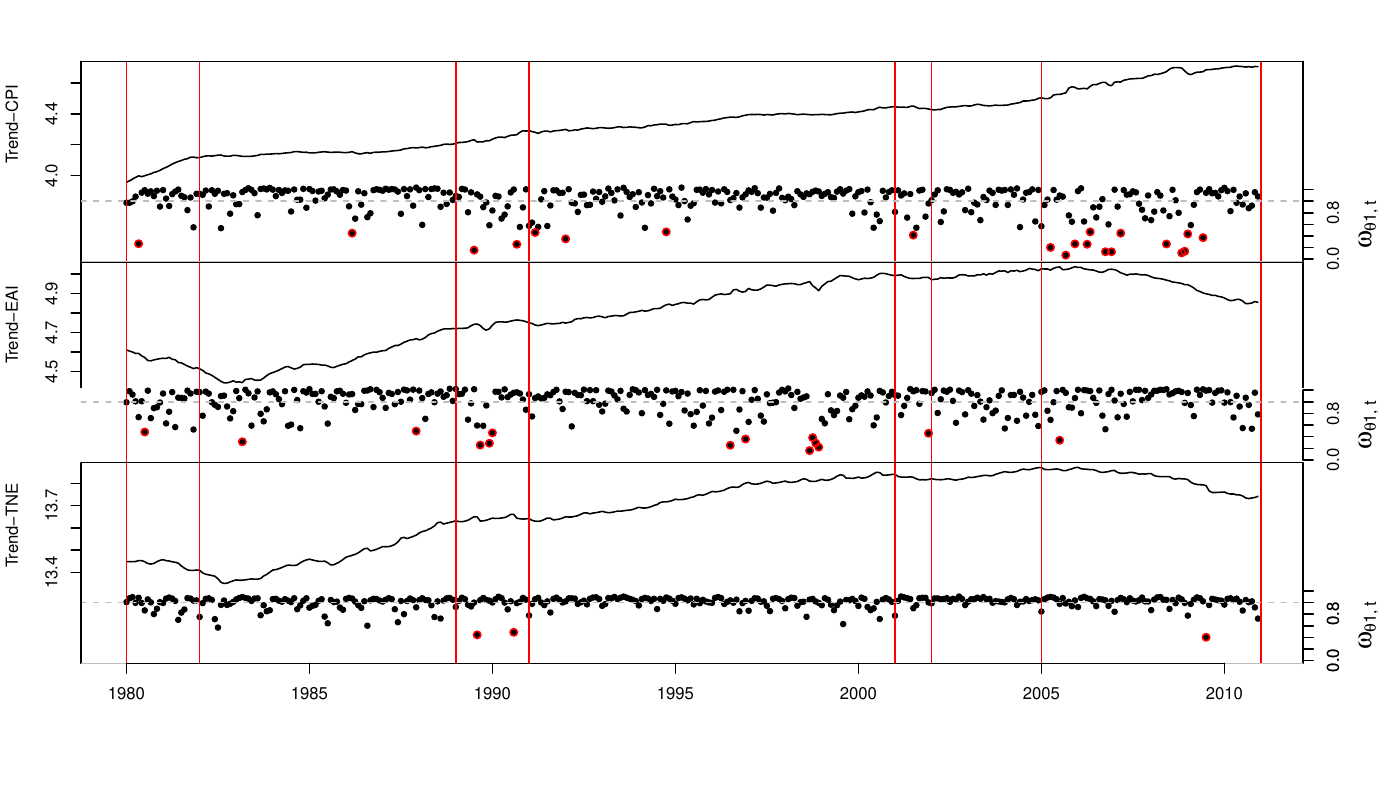}
\caption{Comparison structural breaks for the monthly Consumer
Price and Economic Indexes in Puerto Rico. The right scale
is for the latent $\omega_{\theta,t_{1}}$ parameter. Red points illustrate the outliers and structural breaks.
Red lines divide the recession time periods.} \hspace{-6.5cm}
\end{center}

Figure 6 illustrates the trend, slope and seasonal posterior component with their corresponding credible
intervals for TNE. An interesting feature by using RBDMs is
that the credible intervals are not of constant width. For the seasonal component
the 95\% interval is wider for the structural changes. Figure 7
displays the relationship between the time series for the recession periods.
 The abrupt
changes in both indexes during the last period may be the
consequence of the economic crisis that Puerto Rico has been
suffering since 2006. In the recent economic crisis in Puerto Rico 7816 public employments
lost their jobs in July 2009 and the RBDM detects this change in the bottom of Figure 7.

\end{figure}

\clearpage

\section{Conclusions}

In this paper we apply Robust Bayesian Dynamic Models (RMBDs) to
Latin-American time series from Mexico and Puerto Rico. The classes of RMBDs presented in this work
with weakly-robust priors for the observation and state variances
consider most of the empirical models used in the classical econometrics literature as random walk, linear trend, seasonal and a combination of those models.
We found that using RMBDs allow us
to account historic outliers and structural breaks in the inflation in Mexico and
the economic recession periods in Puerto Rico. In fact, the structural changes have a
contextual historical and economical meaning.  Also, the
model has the feature of producing  not constant credible intervals
over time even after accounting for boundary effects in the Latin-American variables. Finally we consider
that in a future work new RMBDs could be implement to time series with weakly-shrinkage
priors for the observation and state variances as is the case
of generalized autoregressive conditional heteroscedasticity (GARCH) models for dynamic volatility approaches in finance.\\

\textbf{Acknowledgements}\\

We thank to the associated editor and the two referees by detailed
comments and suggestions that greatly improved the quality of this paper.
Also, we thank to Brenda Betancourt (UCSC) and Javier Rubio (Warwick)
for the discussions with the first author which were very useful in preparing the last version of
this paper. Jairo F\'uquene and Marta \'Alvarez were supported for the grant PII-2012, Business School, University of Puerto Rico, Rio Piedras.

\bibliography{biblio_p}

\appendix

\section{Prior distributions and Markov Chain Monte Carlo algorithm}

The scaled
Beta2 distribution can be defined as a scale mixture of Gammas for
the square of the scale as follows:
\begin{align}
\tau^{2}&\sim \text{Gamma}(p,\beta/\rho)\\
\rho&\sim \text{Gamma}(q,1)
\end{align}
where Gamma$(a,b)$ denotes the Gamma distribution:
\begin{equation}
p(x|\alpha,b)=\frac{1}{\Gamma(\alpha)\beta^{\alpha}}x^{\alpha-1}\exp\{-x/\beta\}
\;\;\;\; a>0, b>0,
\end{equation}
with $\beta$ the scale parameter. Therefore the scaled Beta2 prior
for the square scale is the following:
\begin{equation}
\pi(\tau^{2})=\frac{\Gamma(p+q)}{\Gamma(p)\Gamma(q)}\frac{1}{\beta}\frac{\left(\dfrac{\tau^{2}}{\beta}\right)^{p-1}}{\left(1+\dfrac{\tau^{2}}{\beta}\right)^{p+q}}.
\;\;\;\; \tau>0.
\end{equation}
For precisions $\lambda=1/\tau^{2}$, we assign the scaled Beta 2 as
\begin{equation}
\pi(\lambda)=\frac{\Gamma(q+p)}{\Gamma(q)\Gamma(p)}\beta\frac{(\beta\lambda)^{q-1}}{\left(1+\beta\lambda\right)^{p+q}};\;\;\;\;
\lambda>0, \end{equation}
typically the hyper-parameters $p,q$ are fairly small, for example
$p=q=1$, and $\beta$ quite small (see \citeasnoun{pericchi1}) obtaining
a bounded density at the origin, flat tails and a vague prior distribution.\\

The RBDMs can be written in hierarchical form as follows:

\begin{align}
y_{t}&=F_{t}\boldsymbol{\theta}_{t}+\nu_{t} \;\;\;\; \nu_{t}\sim N_{m}(0,V_{t}),\\
\boldsymbol{\theta}_{t}&=G_{t}\boldsymbol{\theta}_{t-1}+\omega_{t}\;\;\;\; \notag
\omega_{t}\sim N_{p}(0,W_{t}),
\end{align}

where the observation and state variances are given by:

\begin{align*}
\hspace{1cm} V_{t}^{-1}=&\lambda_{y}\omega_{y,t}, & W_{t,i}^{-1}=&\lambda_{\theta,i}\omega_{\theta,t_{i}},\\
\lambda_{y}|q& \sim \text{Gamma}(q,(\beta\rho_{y})^{-1}), &
\lambda_{\theta,i}|q& \sim \text{Gamma}(q,(\beta\rho_{\theta,t_{i}})^{-1}),\\
\omega_{y,t}& \sim \text{Gamma}(\upsilon/2,2/\upsilon), &
\omega_{\theta,t_{i}}&
\sim \text{Gamma}(\upsilon/2,2/\upsilon),\\
\rho_{y}&\sim \text{Gamma}(p,1), & \rho_{\theta,t_{i}} &\sim
\text{Gamma}(p,1),
\end{align*}

In order to obtain posterior inference on the state parameters
$\boldsymbol{\theta}_{1}$,...,$\boldsymbol{\theta}_{T}$, we use the forward filtering
backward sampling (FFBS) given in \citeasnoun{Fruwirth} which is
practically a simulation of the smoothing recursions. The FFBS
works as follows:
\begin{enumerate}
\item Use the Kalman Filter equations for (5).
  Let
$m_{0}$ and $C_{0}$ (known) with $(\boldsymbol{\theta}_{0}|D_{0})\sim
N(m_{0},C_{0})$ and
\begin{equation}
\boldsymbol{\theta}_{t}|y_{1:t-1}\sim N(m_{t-1},C_{t-1})
\end{equation}

\begin{itemize}
\item The one step predictive distribution of
  $\boldsymbol{\theta}_{t}$ given $y_{1:t-1}$ is
  Gaussian $(\boldsymbol{\theta}_{t}|D_{t-1})\sim
N(a_{t},R_{t})$ with parameters:
\begin{align}
a_{t}&=G_{t}m_{t-1}; & R_{t}&=G_{t}C_{t-1}G^{'}_{t}.
\end{align}
\item The one step predictive distribution of $y{t}$ given $y_{1:t-1}$ is
  Gaussian $(y_{t}|D_{t-1})\sim
N(f_{t},Q_{t})$ with parameters:
\begin{align*}
f_{t}&=F^{'}_{t}a_{t}; &
Q_{t}&=F_{t}^{'}R_{t}F_{t}+V_{t}.
\end{align*}
\item The filtering distribution of ${\boldsymbol{\theta}}_{t}$ given
  $y_{1:t-1}$ is
  Gaussian $(\boldsymbol{\theta}_{t}|D_{t})\sim
N(m_{t},C_{t})$ with parameters:
\begin{align}
m_{t}&=a_{t}+A_{t}e_{t}; & C_{t}&=R_{t}-A_{t}Q_{t}A^{'}_{t}
\end{align}
where $A_{t}=R_{t}F_{t}Q^{-1}_{t}$, and $e_{t}=y_{t}-f_{t}$.

\end{itemize}
\item At time $t=T$ sample $\boldsymbol{\theta}_{T}$ from
  $N(\boldsymbol{\theta}_{T}|m_{t},C_{t})$.
\item For $t=(T-1):0$ sample $\boldsymbol{\theta}_{t}$ from
  $N(\boldsymbol{\theta}_{t}|m_{t}^{*},C_{t}^{*})$with
\begin{align}\notag
m_{t}^{*}&=m_{t}+B_{t}(\boldsymbol{\theta}_{t+1}-a_{t+1})&
C^{*}_{t}=C_{t}-B_{t}R_{t+1}B^{'}_{t}
\end{align}
where $B_{t}=C_{t}G^{'}_{t+1}R^{-1}_{t+1}$.

\end{enumerate}

In order to obtain now posterior inference for the rest of parameters
in the observation and state variances we use the standard approach
by considering the full conditional
distribution proportional to the joint distribution of all random
variables (parameters) considered. So, for example using (13) the full conditional for $\lambda_{y}$ is given by:

\begin{align}
\pi(\lambda_{y}|...)\propto\prod_{t=1}^{T}\lambda_{y}^{1/2}\exp\left\{-\frac{\lambda_{y}\omega_{y,t}}{2}(y_{t}-F_{t}\boldsymbol{\theta}_{t})^{2}\right\}
\cdot\lambda_{y}^{q-1}\exp\left\{-\beta\rho_{y}\lambda_{y}\right\}, \end{align}

hence,

\begin{align} \lambda_{y}|...\sim \text{Gamma}\left(q+\frac{T}{2},\frac{1}{2}SSy^{*}+\beta\rho_{y}\right) \end{align}

where $SSy^{*}=\sum_{t=1}^{T}\omega_{y,t}(y_{t}-F_{t}\boldsymbol{\theta}_{t})^{2}$. The rest of full
conditional distributions are given by:

\begin{align*} \lambda_{y}|...\sim & \text{Gamma}\left(q+\frac{T}{2},\frac{1}{2}SSy^{*}+\beta\rho_{y}\right), \\
\lambda_{\theta,i}|...\sim&
\text{Gamma}\left(q+\frac{T}{2},\frac{1}{2}SS_{\theta,i}^{*}+\beta\rho_{\theta,t_{i}}\right)& \end{align*}

where $SS_{\theta,i}^{*}=\sum_{t=1}^{T}\omega_{\theta,t_{i}}(\theta_{t_{i}}-(G_{t}\theta_{t-1})_{i})^{2}$ for
$i=1,2,...,p$;

\begin{align*} \omega_{y,t}|...\sim&
\text{Gamma}\left(\frac{\upsilon+1}{2},\frac{\upsilon+\lambda_{y}(y_{t}-F_{t}\boldsymbol{\theta}_{t})^{2}}{2}\right),\\
\omega_{\theta,t_{i}}|...\sim&
\text{Gamma}\left(\frac{\upsilon+1}{2},\frac{\upsilon+\lambda_{y}(\theta_{t_{i}}-\lambda_{\theta,i}(G_{t}\theta_{t-1})_{i})^{2}}{2}\right)
\end{align*}

\begin{align*} \rho_{y}|...\sim&\text{Gamma}\left(p+q,\beta\lambda_{y}+1\right), &
\rho_{\theta,t_{i}}|...\sim&\text{Gamma}\left(p+q,\beta\lambda_{\theta,i}+1\right), \end{align*}

\clearpage

\begin{figure}[h]

\section{Convergence of parameters for the Mexico case}

\begin{center}
\vspace{3.5cm} \hspace{-4.0cm}
\includegraphics[bb=0 0 10cm 10cm,scale=0.8,keepaspectratio]{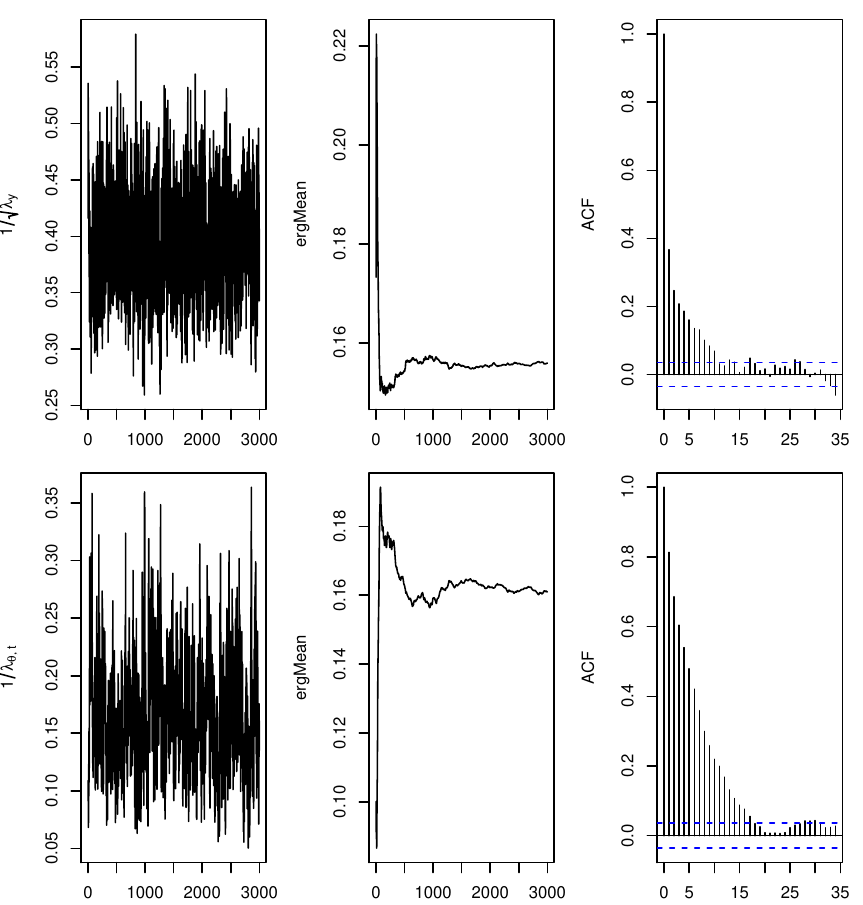}
\caption{Convergence diagnostic plots for the constant precision parameters of CPI-variations from Mexico 1969-1983.
Left: Histograms for the precisions. Middle: Ergodic mean. Right: Autocorrelation plots.}
\hspace{-5.0cm}
\end{center}

\end{figure}

\end{document}